\documentclass{aa}

\usepackage{epsfig}
\usepackage{graphics}
\usepackage{float}
\usepackage{amsmath}
\usepackage{multirow}
\usepackage{longtable}
\usepackage{rotate}
\usepackage{array}
\usepackage{subfigure}
\def\kms{\ifmmode{\rm km\,s^{-1}}\else\hbox{$\rm km\,s^{-1}$}\fi}

\setlongtables

\begin{document}

\title{Spectroscopic binaries with components of similar mass}

\author{L.B.Lucy}
\offprints{L.B.Lucy}

\institute{Astrophysics Group, Blackett Laboratory, Imperial College 
London, Prince Consort Road, London SW7 2AZ, UK}
\date{Received ; Accepted }

\abstract{The assertion that there is an intrinsic excess of binaries 
with mass ratios $q \simeq 1$ - the twin hypothesis - is investigated.
A strong version of this hypothesis (${\cal H}_{s}$), due to
Lucy \& Ricco (1979) and Tokovinin (2000), 
refers to a
narrow peak in the distribution function $\psi(q)$ for $q \ga 0.95$.
A weak version (${\cal H}_{w}$), due to Halbwachs et al. (2003), refers to a
broad peak for $q \ga 0.8$.
Current data on SB2's is analysed and ${\cal H}_{s}$ is found to be 
statistically significant for a sample restricted to orbits of high 
precision. But claims that ${\cal H}_{s}$ is significant for binaries
with special characteristics are not confirmed since the sample sizes are
well below the minimum required for a reliable test.
With regard to ${\cal H}_{w}$, additional observational evidence 
is not presented, but evidence to the contrary in the form of Hogeveen's 
(1992b) model of biased sampling with $\psi \propto q^{-2}$ is criticized.
Specifically, his success in thus fitting catalogued data depends on 
implausible assumptions about the research methodologies of binary-star
spectroscopists.
   }

\authorrunning{Lucy}
\titlerunning{The twin hypothesis}
\maketitle

\section{Introduction}

Statistical studies of binaries are hindered by selection effects.
For example, because of the steep
mass-luminosity relation for main-sequence stars, the discovery of 
double-lined spectroscopic binaries (SB2's) favours systems
with mass ratios $q \simeq 1$. However,
notwithstanding this bias, some authors have
argued that there is an intrinsic excess of binaries with components having
similar masses.
This claim may conveniently be referred to as the twin hypothesis.
Moreover, there are
strong (${\cal H}_{s}$) and weak (${\cal H}_{w}$) versions of this hypothesis.

According to Halbwachs et al. (2003), an excess occurs for $q \ga 0.8$, and 
they refer to such binaries as twins (${\cal H}_{w}$). This claim results from
their finding that $\psi(q)$, the probability density function (pdf) of the
$q$'s of unevolved binaries, has a positive slope for $q \ga 0.75$.
These authors also review earlier determinations
of $\psi(q)$, in particular those claiming a bimodal pdf with one peak
at $q \simeq 1$.     
  
On the other hand, Tokovinin (2000, hereafter Tk00) refers to binaries with
$q \ga 0.95$ as twins (${\cal H}_{s}$). This is motivated by the claims
of Lucy \& Ricco (1979, hereafter LR79) and Tk00 that there is significant
excess of binaries with components having almost identical masses. 
 
Note that  ${\cal H}_{s}$ and 
${\cal H}_{w}$ are not mutually exclusive, nor does one imply the other: either or
both could be true or false.

If true, ${\cal H}_{s}$ offers the tantalizing prospect of a decisive
test of
binary formation theories, which then stand or fall according to  
whether or not they predict a near delta function at $q \simeq 1$. But if
${\cal H}_{s}$ is false, $\psi(q)$ is a smooth and
relatively featureless function, and so 
binary-formation simulations might have to sample the (unknown) distribution
of pre-binary configurations in order to critically test formation theories
An analogy with spectroscopy is
instructive here: 
the discovery of just one weak emission line superposed on a 
featureless continuum dramatically enhances the prospects of theoretical
interpretation.
  
Because of the potential importance of ${\cal H}_{s}$,
this paper adopts a different approach to its investigation from those of
LR79 and Tk00 as well as re-evaluating claims based on small samples. In
addition, Hogeveen's (1992b) finding of a
negatively-sloped $\psi(q)$, which would invalidate both ${\cal H}_{s}$ and
${\cal H}_{w}$, is examined in detail.

Throughout this paper, with the exception of Sect.2.5, binary data is
discussed on the assumption that $\psi(q)$ is independent of binary mass and
angular momentum - see LR79, Sect.I.

\section{The strong twin hypothesis}

In this section, evidence for ${\cal H}_{s}$ is sought from current data on
SB2's, and the significance of this evidence is quantified.

\subsection{Measurement errors}
A major part of LR79 was devoted to correcting the $q$-distribution
of SB2's for measurement errors. The starting point of that investigation was
the conjecture that the preference for $q \simeq 1$ suggested by such
classical binaries as
Y Cyg and YY Gem was less evident for the totality of catalogued SB2's
because of the inclusion of less precise orbits. Accordingly, LR79
went back to the discovery papers and derived or computed the standard
errors $\sigma_{q}$. This excercise revealed that 
systems with $\sigma_{q} \ga 0.05$ were not uncommon, and so any narrow peak
at  $q \simeq 1$ would inevitably be softened by measurement errors.
The error-free pdf was then estimated using a version of the Richardson-Lucy
(R-L) deconvolution technique for data sets where each measurement $q$ has a
different 
$\sigma_{q}$. The resulting $\psi(q)$ has a sharp peak at
$q \simeq 0.97$, accounting for about half of binaries in the
interval (0.95,1.0) - see Fig. 3 in LR79. 

The use of the R-L technique by LR79 was later criticized by Hogeveen (1992a).
He argues that the peak at $q \simeq 0.97$ is an
artefact of the deconvolution procedure. But Tk00's confirmation of
${\cal H}_{s}$
cannot be similarly challenged since it does not rely on
deconvolution.

\subsection{SB2 sample from the 9th catalogue}

The SB2's investigated in LR79 were drawn from the 6th 
catalogue of the orbital elements of spectroscopic binary systems.
The current greatly expanded version is the 9th catalogue
(Pourbaix et al. 2004), and this will be the source of data for this
investigation - specifically, the updated online version as at 15.04.05. 

In order to re-investigate the effect of errors on the $q$-distribution, 
all SB2's were selected for which the components' semi-amplitudes
$K_{1,2}$ and their standard errors $\sigma_{K_{1,2}}$ are recorded.
Then, with $q$ defined to be $<1$, all systems with $q > 0.84 $
were selected, and their $\sigma_{q}$'s computed from 
the propagation-of-errors formula,
\begin{equation}
  \left(\frac{\sigma_{q}}{q} \right)^{2} =
  \left( \frac{\sigma_{K_{1}}}{K_{1}} \right)^{2}
                       + \left( \frac{\sigma_{K_{2}}}{K_{2}} \right)^{2}
\end{equation}
The result is a sample of 211 systems, with $\sigma_{q}$ ranging from
0.0012 to 0.19. A histogram showing
the distribution of the $\sigma_{q}$'s is given in Fig.1.

\begin{figure}
\vspace{8.2cm}
\includegraphics{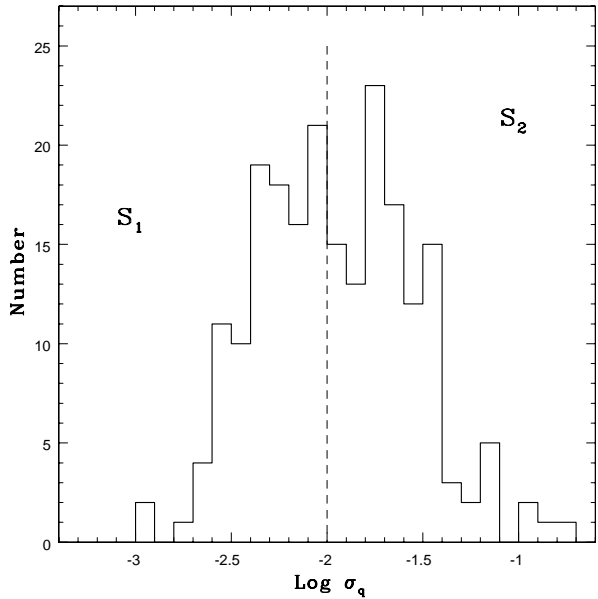}
\caption{Distribution of standard errors $\sigma_{q}$ for 211 SB2's with
$q > 0.84$. The
vertical dashed line divides the data into high- and
low-precision samples, labelled $S_{1}$ and $S_{2}$, respectively.}
\end{figure}

   This data is now divided into a high precision sample ($S_{1}$) comprising
102 systems with  $\sigma_{q} < 0.01$ and a low precision
sample ($S_{2}$) comprising 109
systems with  $\sigma_{q} >  0.01$. For each of these samples,
the $q$'s are counted into overlapping bins of width 0.02 with
centres separated by 0.01. The resulting counts $N$ together with  
$\sqrt{N}$ error bars are plotted in Fig. 2.

\begin{figure}
\vspace{8.2cm}
\includegraphics{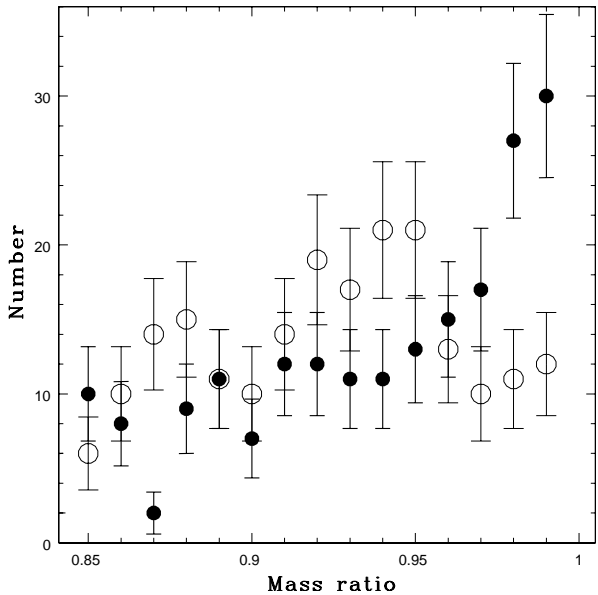}
\caption{Mass ratio distributions for high- and low-precision samples. The
filled circles are for 102 binaries with
$\sigma_{q} < 0.01$, and the open circles for 109 systems with 
$\sigma_{q} > 0.01$. Error bars are $\pm \sqrt{N}$.}
\end{figure}

   Fig 2 supports the basic premise of
LR79 that there is a narrow peak at $q \simeq 1$ that is   
partially obscured by measurement errors.
With $\sigma_{q} < 0.01$, negligible error-broadening
occurs for $S_{1}$, and the plot shows marked excesses in the bins
centred at 0.99 and 0.98 - i.e., for the interval (0.97,1.0). In contrast,
for $S_{2}$, where significant broadening is inevitable, no peak is
evident.   

   Note that this dependence of the $q$-distribution on precision, as
posited by LR79, is absent from the simulations of Hogeveen (1992a, Sect.5).
Instead of deriving the $\sigma_{q}$'s from the discovery papers as in LR79
and here, he assigns random artificial errors $\sigma_{q}$ and so
effectively his $q$-distributions are impervious to degradation by
measurement errors.
His simulations therefore miss
the essential effect of the high information content from high
precision
orbits being imposed on the less precise data. This same effect is striking
when the R-L algorithm is applied to the problem of co-adding images
with different psf's (Lucy 1991). 

   The peak at $q \simeq 1$ for $S_{1}$ indicates, in agreement with Tk00,
that 
${\cal H}_{s}$ need not now rely for its support on a deconvolution procedure
as in LR79.

\subsection{Tests of significance}

The error bars in Fig. 2 give a rough indication of the significance
of the differences between $S_{1}$ and $S_{2}$. To quantify
this significance, we now test the null hypothesis $H_{0}$ that the two
samples are drawn form the same parent population.

The simplest test is to construct a $2 \times 2$ contingency table. To do
this, the two samples ($i = 1,2$) are themselves divided at $q = q_{d}$ into
low- and high-$q$ subsamples, labelled $j=1$ and $j=2$, respectively.

Taking the hint provided by Fig 2, we take $q_{d} = 0.97$. The resulting counts
are $N_{11} = 62$,  $N_{12} = 40$, $N_{21} = 93$,$N_{22} = 16$. The expected
values on $H_{0}$ are: 
$n_{11} = 74.9$,  $n_{12} = 27.1$, $n_{21} = 80.1$, $n_{22} = 28.9$,
resulting in $\chi^{2} = 16.3$ with 1 degree of freedom - see, e.g.,
Press et al. (1992, p.624). The probability on $H_{0}$ that
$\chi^{2} > 16.3$ is $p = 5.5 \times 10^{-5}$. Accordingly, we reject
$H_{0}$   
and conclude that the differences between $S_{1}$ and $S_{2}$ are
highly significant. 

This test can be repeated with different values of $q_{d}$. With
$q_{d} = 0.95$, $\chi^{2} = 14.2$, corresponding to $p = 1.6 \times 10^{-4}$,
a negligible loss of significance.
But with $q_{d} = 0.94$ or lower, the peak is increasingly diluted by the
background, and the significance of the rejection of $H_{0}$ drops sharply.
This indicates that the lower limit of the peak associated with ${\cal H}_{s}$
occurs at $q \simeq 0.95$, in agreement with Fig. 3 in LR79.

Another approach to quantifying the significance of differences between 
$S_{1}$ and $S_{2}$ is the Kolmogorov-Smirnov (K-S) test, which has the
merit of not requiring  
binning. But here, because the differences are at extreme end of the
range - i.e., at $q \simeq 1$, we follow Press et al. (1992, p.620) and
use the K-S variant due to N.H.Kuiper. Applying the Kuiper test to the ordered
$q$'s in $S_{1}$ and $S_{2}$, we find that Kuiper's statistic
$V = D_{+} + D_{-} = 0.33$, and the probability on $H_{0}$ that $V > 0.33$
is $p = 2.7 \times 10^{-4}$, confirming highly significant differences.

Because Kuiper's variant is not as familiar as the original K-S test,
a Monte Carlo version has also been carried out. With $m=0$ initially, 
the following
simulation is carried out $M$ times : from the 211 values of $q$,
102 are randomly selected for assignment to $\tilde{S}_{1}$, with the
remainder assigned to $\tilde{S}_{2}$.
The Kuiper statistic $V$ is then
computed from the ordered $q$'s in $\tilde{S}_{1}$ and $\tilde{S}_{2}$,
and we set $m = m+1$ if $V > 0.33$. Now, by construction,
$\tilde{S}_{1}$ and $\tilde{S}_{2}$ are drawn from the same parent
population, and so $H_{0}$ is true. Accordingly, at the conclusion of the
experiment, $m/M$ is an unbiased estimate of the probability
on $H_{0}$ that $V > 0.33$. An experiment with $M = 10^{7}$ gives
$m = 2302$, so that $p = 2.3 \times 10^{-4}$, in satisfactory agreement 
with the estimate given by Kuiper's asymptotic theory.

\subsection{Minimum sample sizes}

The above tests show that a sample of 102 binaries with $q > 0.84$ and
$\sigma_{q} < 0.01$ suffices to claim the discovery of a significant peak
at $q \simeq 1$ and thus support ${\cal H}_{s}$. But if $N$, the number
of binaries with accurate orbits, were reduced, the peak's significance would
drop and the support for ${\cal H}_{s}$ would weaken. Thus, as $N$
decreases, there is an increasing chance of a false negative - i.e. of
rejecting ${\cal H}_{s}$ even though it is true. 

This risk can be crudely quantified with a bootstrap technique.
The following calulation is repeated numerous times
as a function of $N$: a sample $\tilde{S}_{1}$ of $N$ binaries with
$\sigma_{q} < 0.01$ is created,
each binary being randomly selected from $S_{1}$ - i.e., sampling with
replacement. Kuiper's test is then applied
to compare $\tilde{S}_{1}$ with $S_{2}$, and the probability $p$ 
computed that the Kuiper statistic could be exceeded if $H_{0}$ is true.    
From these, we can derive the percentages
of the samples $\tilde{S}_{1}$ for which $H_{0}$ is {\em not} rejected at the
1 and 5 percent levels of significance. These error rates are plotted as a
function of $N$ in Fig. 3.

\begin{figure}
\vspace{8.2cm}
\includegraphics{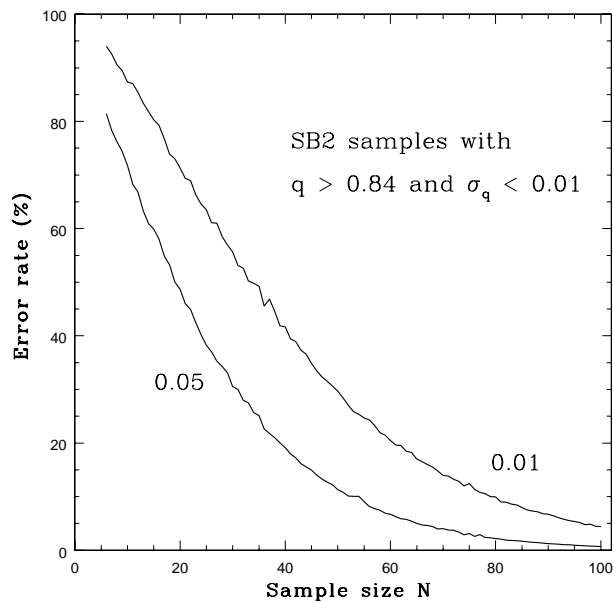}
\caption{Error rates as function of sample size $N$ when testing
${\cal H}_{s}$. The samples comprise binaries with $q > 0.84$ and
$\sigma_{q} < 0.01$ and are constructed by bootstrap resampling from the
102 binaries in $S_{1}$. The percentage rates are plotted for the 1 and 5
percent levels of significance.}
\end{figure}

A minimum sample size $N_{min}$ can be defined as being such that for
$N < N_{min}$ there is a $> 10$ percent risk of failing to confirm
${\cal H}_{s}$ even though it is true. From Fig. 3, we find that 
$N_{min} \simeq 79$ and $\simeq 54$ for the 1 and 5 percent levels 
of significance, respectively. These results confirm and quantify the
remark in Tk00 that ${\cal H}_{s}$ cannot be detected in small samples.

\subsection{Limiting period}

From the data then available, LR79 concluded that ${\cal H}_{s}$ is no
longer significant for binary periods $P > 25$ days. The existence of a
cut-off period $P_{L}$ was confirmed by Tk00 and Halbwachs et al (2003) but
revised upwards to 40 and 50 days, repectively.

The Kuiper test can also be used to estimate $P_{L}$ from the samples
$S_{1}$ and $S_{2}$. The calculation proceeds as follows: a sample  
$\tilde{S}_{1}$ is created from $S_{1}$ by excluding binaries with $P > P_{*}$.
Kuiper's test is then applied to compare $\tilde{S}_{1}$ and $S_{2}$ and
the probability $p$ computed that these samples are drawn from the same
parent population.

Such calculations show that excluding long period systems indeed increases
the significance of ${\cal H}_{s}$. A plot of $p$ against $P_{*}$ shows
a broad minimum with $p \sim 2.5 \times 10^{-5}$ for $P_{*}$ in the
interval $(24,43)$ days, with $p$ increasing sharply for longer periods.
Thus $P_{L} \simeq 43$ days is the best estimate of the period above which
the data markedly dilutes the significance of ${\cal H}_{s}$. To within errors, this agrees with the estimates of Tk00 and Halbwachs et al. (2003). No doubt,
with increased data, this cut-off period will be found to depend on binary
mass.

\section{Special samples}

In this section, we evaluate claims that ${\cal H}_{s}$ receives support from
binaries with special characteristics. 

\subsection{BY Draconis binaries}

In LR79, data for 9 binaries among the 17 BY Dra variables known at that time
was cited as providing strong support for ${\cal H}_{s}$. Subsequently,
at the suggestion of an anonymous referee, Fekel et al. (1988) re-examined
the evidence for a peak at $q \simeq 1$ on the basis of an enlarged sample
of 20 BY Dra binaries. It turns out that the additional binaries fill in
the interval (0.80,0.95), resulting in a roughly uniform distribution from
0.80 to 1.0, and so Fekel et al. (1988) correctly report that there is
no longer evidence for a sharp peak at $q \simeq 1$. Nevertheless, in
view of the small sample size, it does not
follow that ${\cal H}_{s}$ must be rejected for the BY Dra binaries.

Of the 20 binaries in the Fekel et al. sample, 8 fall in the interval
(0.84, 1.0) and only 3 have $\sigma_{q} < 0.01$. But even if all the orbits
were precise, the sample of 8 is well below the minima derived in 
Sect. 2.4, and so there is a considerable risk of rejecting
${\cal H}_{s}$ even if it is true. Specifically, for $N = 8$, Fig.3 tells us
that there is a 76 percent chance of  
failing to confirm ${\cal H}_{s}$ at the 5 percent level and a 91
percent chance at the 1 percent level.

\subsection{Eclipsing binaries in the SMC}

Recently, Pinsonneault \& Stanek (2006) have claimed support for
${\cal H}_{s}$ from a sample of 50 eclipsing SB2's of spectral types O and
B in the Small Magellanic
Cloud (SMC). The spectroscopic observations were obtained by
Harries et al. (2003) and Hilditch et al. (2005) who describe their radial
velocities as being of modest quality, no doubt reflecting practical
difficulties of their pioneering project. Of these 50 systems, 15 have
$q$'s in the interval (0.84,1.0), and these
have $\sigma_{q}$'s ranging from 0.013 to 0.12. Thus, compared to
the Galactic SB2's discussed in Sect.2, these orbits would be assigned to the
low precision sample. 

Even if these 15 SMC binaries had precise orbits, the sample size is far below
the minima derived in Sect. 2.4. Nevertheless, in view of the strong claim by
Pinsonneault \& Stanek, the SMC sample has been compared to S2 to see if
$H_{0}$ is rejected. The resulting Kuiper statistic is $V = 1.24$ and the
probability on $H_{0}$ that $V > 1.24$ is $p = 0.48$. Thus the claim that
${\cal H}_{s}$ is supported by the SMC binaries is not even a $1\sigma$
result.

\section{Selection effects}

Before concluding that the narrow peak at $q \simeq 1$ is astrophysically,
as distinct
from merely statistically significant, we must show that it has not been
imprinted on the data by selection effects. In Sect.VI of LR79, five
selection effects were discussed in detail and found not to be capable of
sharply reducing the discovery probability already at $q \simeq 0.95$, the
shortward edge of the narrow peak. Three further selection effects are now
similarly discussed.

\subsection{Biased sampling}

Hogeveen (1992b) has argued that the positive slope of $\tilde{\phi}(q)$,
the $q$-distribution of catalogued SB2's, is entirely due to selection effects.
More specifically, by constructing a synthetic SB catalogue, he claims to
demonstrate with a model of biased sampling
that the montonically {\em decreasing} pdf
\begin{equation}
  \psi(q) \propto q^{-2}  \;\;\;  for \;\;\;      q > 0.3
\end{equation}
that he derives for SB1's is valid also for SB2's - i.e., for $q \ga 0.65$.

To understand and evaluate this remarkable claim, his work 
is here repeated, the only change being a less abstract terminology.
The basic assumption is that Galactic binaries have the $q$-distribution given
by Eq. (2). Hogeveen's model for predicting the contents of the SB catalogue
is then the following: 
the solar neighbourhood is divided into non-overlapping volumes $V_{i}$,
with observer $i$ being awarded exclusive rights to $V_{i}$.
Each observer's task is to discover, monitor and
eventually publish orbits for some or all of the $n_{i}$ binaries in
his assigned search volume.
The synthetic catalogue then comprises the ensemble of published orbits.

In implementing this model, Hogeveen (1992b, Appendix B) assumes that,
independently of $n_{i}$, observer $i$ 
publishes just one orbit, that of the system closest to $q = 1$.
At one extreme, this happens when only one binary is discovered
{\em and} the search procedure
stongly favours $q \simeq 1$. At the other extreme, all
$n_{i}$ binaries are discovered but observer $i$ then selects,
as 'most suitable' 
for monitoring and publication, the system closest to $q = 1$.

\begin{figure}
\vspace{8.2cm}
\includegraphics{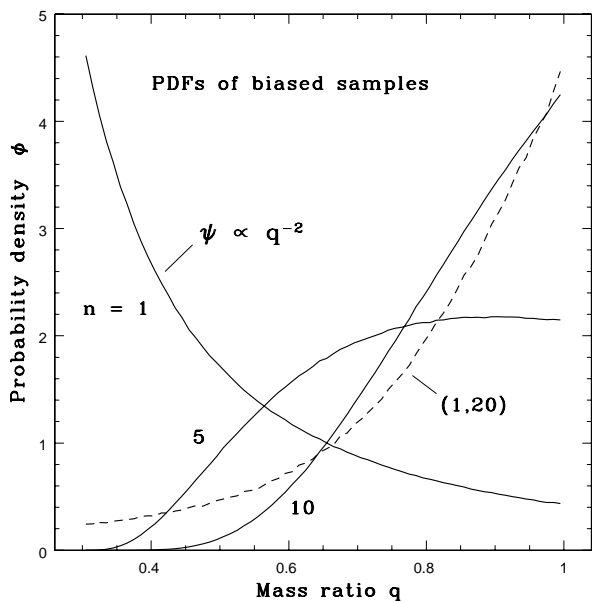}
\caption{Probability density functions $\phi(q)$ of biased samples according
to the model of Hogeveen (1992b). The values of the parameter $n$ are
indicated. For $n = 1$, the true pdf is recovered, namely
$\psi \propto q^{-2}$. The dashed curve is Hogeveen's preferred solution
which has the $n_{i}$ distributed uniformly in the interval (1,20). }
\end{figure}

Fig.4 shows the resulting biased pdf's $\phi(q)$ of synthetic catalogues.
These
simulations divide space into $10^{6}$ volumes $V_{i}$ so that sampling errors
are negligible.
The simplest version of the
model sets $n_{i} = n$, a constant. For $n = 1$, all binaries are discovered
and published so that $\phi_{1}(q) = \psi(q)$, the true distribution. But 
$\phi_{1}(q)$ is monotonically decreasing and so is in conflict with the
positive slope of $\tilde{\phi}(q)$. Fortunately, for $n > 1$,
the assumed
bias towards $q = 1$ takes effect, and $\phi_{n}(q)$ becomes
monotonically increasing for $n \ga 6$. Clearly, as $n \rightarrow \infty$,
each observer's published orbit has $q \rightarrow 1$, and so $\phi_{n}(q)$
tends to a delta function at $q = 1$.  

Because the slope of $\tilde{\phi}(q)$ - see Fig.8 in Hogeveen (1992b) -
is bracketed by the slopes of $\phi_{1}(q)$ and $\phi_{\infty}(q)$, there is
some intermediate $n \simeq 10$ that roughly fits the data. Moreover, this fit
can be improved by exploiting the freedom to depart from constant $n_{i}$.
Specifically, with the $n_{i}$
uniformly distributed in the interval $(1,20)$, Hogeveen achieves a fit to 
$\tilde{\phi}(q)$ that is accepted by the K-S test.   

Given this success, Hogeveen's (1992b) work appears
to stongly support the pdf given in Eq.(2); consequently,
several groups
(Gualandris et al 2005; Ciardullo et al. 2005; Dionne \& Robert 2006)
have adopted this pdf in their population synthesis codes.
Yet others (Pols \& Marinus 1994; Portegies \& Verbunt 1996)
cite Hogeveen but adopt the somewhat less steeply declining pdf
$\psi(q) = 2/(1+q)^{2}$ suggested by Kuiper (1935). These analytic pdf's
are compared in Fig.5 to the piecewise-linear model of Halbwachs et al.
(2003). To avoid complexity, the approximately flat pdf for $q > 0.3$
found by Mazeh et al. (2003) is not included.

\begin{figure}
\vspace{8.2cm}
\includegraphics{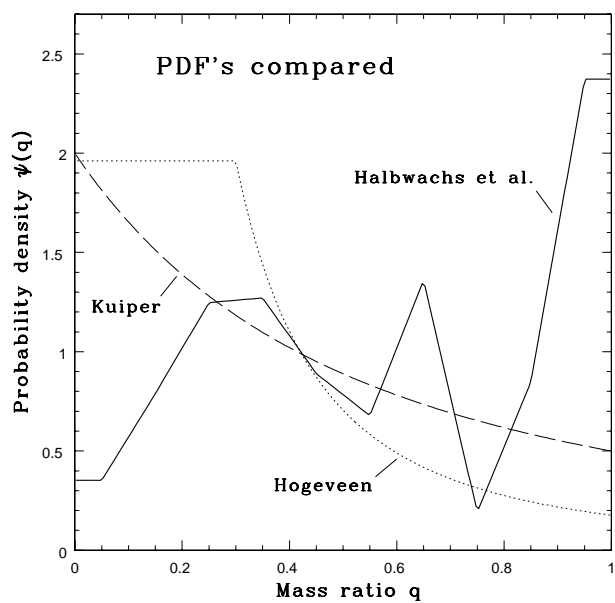}
\caption{Comparison of pdf's. The analytic formulae $\psi(q)$ proposed
by Kuiper(1935) and Hogeveen (1992b) are compared with the corrected
q-distribution given in Fig.7(a) of Halbwachs et al. (2003).}
\end{figure}

However, the K-S test does not provide compelling support for
Hogeveen's $\psi(q)$: the sensitivity illustrated in Fig.4 of $\phi(q)$
to the assumptions about $n_{i}$ all but guarantees that a good fit
can be achieved. Accordingly, this success should not of itself induce belief
in Eq.(2), especially not if the model and its fitted parameters 
are an implausible description of how spectroscopists conduct themselves.

The following comments on the basic elements of Hogeveen's model suggest that
it is indeed implausible:
\indent
    1) Of the $n_{i}$ binaries in $V_{i}$, only the system closest to $q=1$
has its orbit published.

Hogeveen justifies this bold assumption by supposing that an observer,
having discovered several SB2's, selects, for monitoring and publication,
the system whose components have the most 
nearly equal line depths since an
accurate orbit is then more readily achieved. But spectroscopists,
being aware of the extra information obtainable from SB2's, surely do not
fail in this way to capitalize on their own discoveries. Moreover, they do
not set aside difficult systems, as solutions for triple and quadruple systems
demonstrate. 

Alternatively, Hogeveen supposes that systems just above the detection
threshold for SB2's are
underrepresented in the SB catalogue and correspondingly overrepresented
among observers' as yet unreported discoveries.
He conjectures that such binaries remain longer on observing programs so that
acceptable accuracy can be achieved for $K_{2}$.
Fig.6 tests this conjecture by plotting $N_{2}$, the number of velocities
measured for the less massive component, against $q$ for systems in the
9th catalogue. There is no evidence that observers accumulate more data for
systems with $q \la 0.7$.

\begin{figure}
\vspace{8.2cm}
\includegraphics{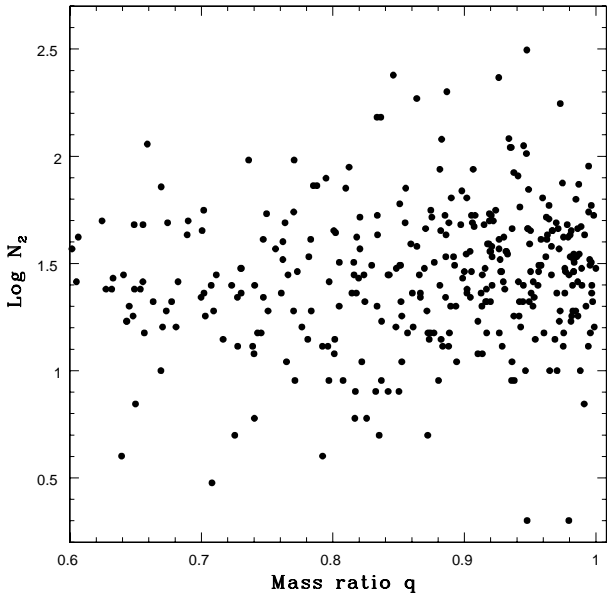}
\caption{Scatter plot of $N_{2}$, the number of radial velocities measured for
the less massive component, against $q$, the mass ratio. All binaries 
from the 9th catalogue (Pourbaix et al. 2004) with $q > 0.6$ are plotted.}
\end{figure}
\indent
    2) The $n_{i}-1$ binaries in $V_{i}$ not monitored and published by
observer $i$ remain unobserved and are thus excluded from the SB catalogue.

This aspect of 
Hogeveen's model follows from granting observer $i$ exclusive rights to
$V_{i}$.
But, in reality, there is no such exclusivity, and so an inefficient observer
$(1/n_{i} \ll 1)$ would often be scooped by competitors
observing the same patch of sky. Accordingly, Hogeveen's reliance on
such observers to achieve a positively-sloped
$\phi(q)$ is especially dubious.

\begin{figure}
\vspace{8.2cm}
\includegraphics{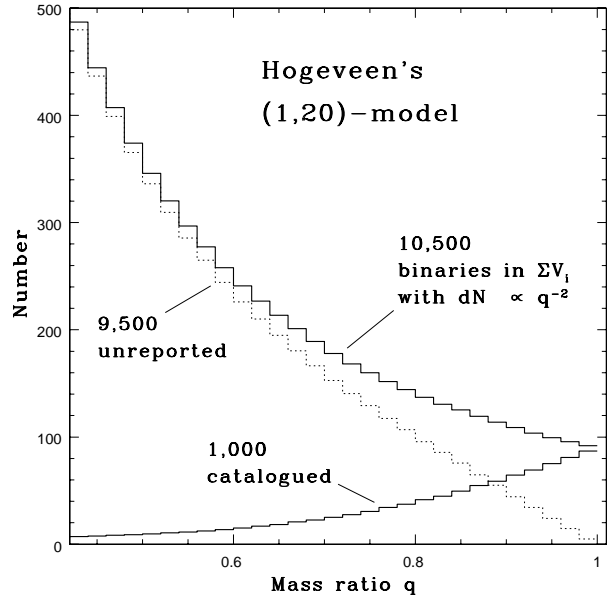}
\caption{Histograms of $q$-distributions for Hogeveen's (1992b) preferred
model, normalized so that catalogued SB's number 1000. The corresponding
total number of binaries in the volumes $V_{i}$ is 10,500 and so the number
of undiscovered or unreported systems is 9,500.}
\end{figure}

The extent to which Hogeveen's fit to $\tilde{\phi}(q)$ depends on
excluded binaries is worth noting, as is their $q$-distribution.
With $n_{i}$ uniformly distributed in (1,20), the average number of binaries
in a search volume is 10.5. Accordingly, if there are 1000 
$V_{i}$, then, out of a total of 10,500 binaries in $\Sigma V_{i}$,
the orbits of 1000 are catalogued, leaving 9,500 undiscovered or unreported.
The $q$-distributions of these three categories are plotted in Fig.7.
We see that, with decreasing $q$, Hogeveen's preferred model implies
a steep decline in discovery
probability or of spectroscopists' inclination to publish.

Note that 7,750 binaries out of the total of 10,500 are in
volumes assigned exclusively to inefficient observers with
$1/n_{i} < 0.1$ whose activities result in a mere 500 catalogued orbits.
It is surely not true that such a paucity of effort by spectroscopists 
applies to 73.8 per cent of the binaries in the solar
neighbourhood.

Hogeveen's (1,20)-model is an attempt to simulate the haphazard manner 
in which the SB catalogue is assembled. As such, it has no 
relevance to the organized and disciplined multi-year surveys that underpin
the work of Halbwachs et al. (2003). Accordingly, there is no possibility
of invoking biased sampling to overturn their evidence in favour of
${\cal H}_{w}$ nor, in particular, to achieve consistency with Eq.(2).

The above analysis of Hogeveen's (1992b) model indicates that his treatment of
biased sampling does not 
provide convincing support for Eq.(2). Indeed, the extremes necessary to
achieve a fit to $\tilde{\phi}(q)$ suggest that this pdf
is strongly excluded.
In any case, his work should not be interpreted as falsifying either
${\cal H}_{w}$ or ${\cal H}_{s}$.

\subsection{Eclipse-probability bias}

In Sect. 3.2, we saw that the sample of SMC binaries is too small
and the $q$'s too imprecise to support ${\cal H}_{s}$. But the possibility
remains that
the data might support ${\cal H}_{w}$, since precision is less of an issue in
confirming a broad peak with FWHM $\sim 0.13$ - see Fig.5.

Binaries in the SMC sample selected themselves by exhibiting eclipses.
This sample is therefore biased in favour of high eclipse probabilities,
and there is no reason to suppose that a sample thus biased will be
unbiased with respect
to $q$.  
Accordingly, $\psi(q)$ cannot be determined by simply binning the
$q$'s of these eB's.

However, the recent paper by S\"{o}derhjelm \& Dischler (2005) shows how one
might proceed in estimating $\psi(q)$. They have developed a population
synthesis code 
that predicts the properties of eB's as a function of eclipse
depth and orbital period.
With the appropriate selection criteria imposed,
the $q$-distribution $\tilde{\phi}(q)$ of the SMC eB's    
could be predicted with this code.
The input $\psi(q)$ could then be varied
until a satisfactory fit is obtained to $\tilde{\phi}(q)$.

Although in principle possible, this approach is not practicable.
In order to extract $\psi(q)$, we would have to input 
other statistical properties of SMC binaries, specifically the distributions
of their separations and eccentricities. But these are not well determined
even for Galactic binaries.

From Table A.2 of S\"{o}derhjelm \& Dischler (2005), the fraction
of OB binaries showing detectable eclipses is $\sim 10^{-3}$. To 
determine statistical properties reliably from such a small
fraction requires intelligently-contolled sampling , as in
opinion polling, or sampling with accurately
known bias. Neither of these criteria obtain for the SMC eB's.   

The above remarks demonstrate that this sample of eB's is
not suitable for deriving the statistical properties of binaries in the 
SMC. But such was not among the aims of 
Harries et al. (2003) and Hilditch et al. (2005). Their focus is on
fundamental parameters in order to investigate single- and binary-star
evolution at low metallicity and to derive the distance to the
SMC.    
Nevertheless, their successful application of a multi-object spectrograph
to obtain numerous SB orbits in the SMC indicate that a survey of
non-eclipsing massive stars is not infeasible. By discovering and then
monitoring
SB's in the SMC (or LMC), such a survey could yield useful information
on binary statistics in an extragalactic population.   

Although directed at the SMC sample, these negative arguments 
apply to any sample of eB's. This explains why statistical
studies of $\psi(q)$ have not hitherto been based on eB's. The only
viable alternative
to SB samples is the distribution of offsets in colour-magnitude diagrams due
to binarity. K\"{a}hler (1999) has analysed such photometric binaries   
in the Pleiades and finds a bimodal $q$-distribution with a broad peak at
$q \simeq 1$, thus supporting
${\cal H}_{w}$. The modest number of offsets and the limited precision of
their measurement preclude a decisive test of ${\cal H}_{s}$. 

\subsection{Precision bias}

In order to avoid deconvolution, a high-precision sample has been used in
Sect.2 to
confirm ${\cal H}_{s}$. But,
as the referee (J.L.Halbwachs) points out, this selection itself introduces
a bias
that to some degree favours $q \simeq 1$.
For an SB2 with $q \neq 1$, the lines of the secondary are generally the
weaker,
and so we expect that $\sigma_{K_2}/K_{2} > \sigma_{K_1}/K_{1}$.
Moreover, this disparity
will increase with decreasing $q$ thus reducing the fraction of SB2's
meeting a fixed precision criterion such as $\sigma_{q} < 0.01$. 

This effect can be investigated for SB2's in the
9th catalogue. From Eq.(1), the fractional contribution
of the secondary ($K_{2} > K_{1}$) to the variance $\sigma_{q}^{2}$ is 
\begin{equation}
  f_{2} = \left( \frac{\sigma_{K_{2}}}{K_{2}} \right)^{2}/ \left[
  \left( \frac{\sigma_{K_{1}}}{K_{1}} \right)^{2}
                       + \left( \frac{\sigma_{K_{2}}}{K_{2}} \right)^{2}  
                       \right]
\end{equation}
and we may then readily show that
\begin{equation}
 \sigma_{q} = \sigma_{q}^{*} / \sqrt{2(1-f_{2})}
\end{equation}
where $\sigma_{q}^{*}$ is the standard error of $q$ when
$\sigma_{K_2}/K_{2} = \sigma_{K_1}/K_{1}$.

The variation of $f_{2}$ with $q$ is shown in Fig.8, where the binning is
the same as for Fig.2 but extending to smaller $q$. For each bin, the mean
$\bar{f_{2}}$ is plotted, with error bars indicating its standard error.
From this figure, we see that for $q \ga 0.9$, there is a barely significant
departure from $\bar{f_{2}} = 0.5$, the line along which both components
contribute equally to $\sigma_{q}^{2}$. But for $q \la 0.9$, the departure
is significant and in the expected direction. Thus, for $q \sim 0.8$, 
secondaries contribute $\sim 70$ per cent of the variance 
$\sigma_{q}^{2}$, corresponding to $\sigma_{q}/\sigma_{q}^{*} \sim 1.3$.  

The implications of this excercise are as follows: if $\psi(q)$ were being
determined over an extended interval of $q$ from a precision-limited sample,
then this bias would be of concern. However, since this effect is
inconsequential for $q \ga 0.9$, precision-bias cannot account for 
the peak at $q > 0.95$ in Fig.2.

In LR79, several selection effects were
eliminated as explanations of the $q > 0.95$ peak on the basis of theoretical
modelling. This discussion of precision bias is notable as being purely
empirical.

\begin{figure}
\vspace{8.2cm}
\includegraphics{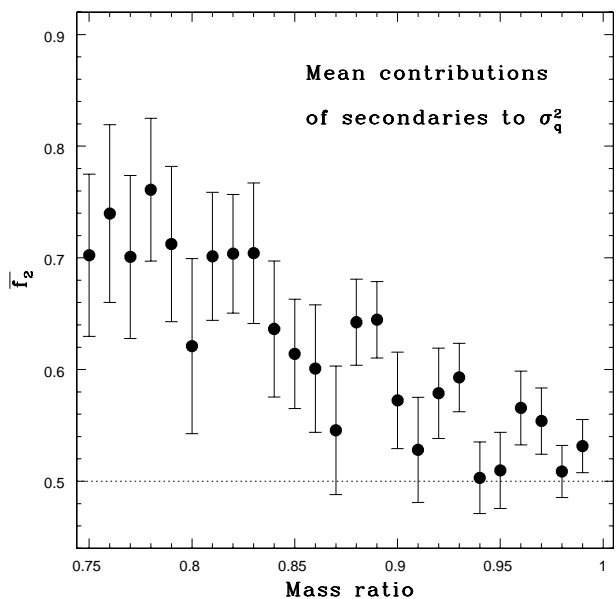}
\caption{Mean fractional contributions $\bar{f_{2}}$ of secondaries to the
variance $\sigma_{q}^{2}$ as function of mass ratio $q$. The error bars denote
the standard errors of the estimates $\bar{f_{2}}$.  Primary and secondary
contribute equally along the dotted line $\bar{f_{2}} = 0.5$  .}
\end{figure}
\section{Conclusion}
The aim of this paper has been to define and investigate the weak and
strong versions of the twin hypothesis for binary stars. For ${\cal H}_{w}$,
no evidence has been offered beyond that of Halbwachs et al. (2003).
Nevertheless, the case for ${\cal H}_{w}$ has been greatly strengthened
by the re-investigation in  Sect.4.1 of Hogeveen's (1992b) model of biased
sampling. Although Hogeveen's work serves as a useful warning of how
misleading inhomogeneous catalogues can be, his success in
fitting observational data requires such extreme assumptions about how
spectroscopists carry out their research programmes that
$\psi(q) \propto q^{-2}$ is surely excluded.
Accordingly,
population synthesis investigations based on the $\psi(q)$ of Hogeveen or
Kuiper should be re-interpreted as exploring the consequences of an 
unrealistic depletion of binaries with $q \simeq 1$.

With regard to ${\cal H}_{s}$, a narrow peak comprising
systems with $q \ga 0.95$ is found to be highly significant, confirming
the work of LR79
and Tk00. Moreover, this peak is now evident without recourse to deconvolution
procedures. However, it remains true that support for ${\cal H}_{s}$  
derives from analyses of catalogued systems and not from a targeted survey
as for  
${\cal H}_{w}$. Unfortunately, it is a major undertaking to carry out an
independent survey large enough to meet the minimum sample sizes estimated
in Sect. 2.4. More feasible would be a programme to re-observe SB2's
in the low-precision sample $S_{2}$ of Sect.2.2 in order to reduce
$\sigma_{q}$ to $< 0.01$. If ${\cal H}_{s}$ is true, the narrow peak should
increase in significance as orbits of low precision are thus upgraded. This
is a project where orbits obtained with small and medium-sized telescopes
are a worthwhile contribution (cf. Tomkin \& Fekel 2006).

\begin{acknowledgements}
I am grateful to F.C.Fekel, R.F.Griffin and C.D.Scarfe for responding promptly
and in detail to queries related to this investigation and to the referee,
J.L.Halbwachs, for a thought-provoking report.
\end{acknowledgements}

\end{document}